\documentclass[aps,prd,english,10pt,preprintnumbers,amsmath,amssymb,nofootinbib,twocolumn,superscriptaddress]{revtex4-1}

\pdfoutput=1

\usepackage[utf8]{inputenc}
\usepackage{graphicx}
\usepackage{bbm}
\usepackage{amssymb}
\usepackage{amsmath}
\usepackage{tabularx}
\usepackage{color}
\usepackage{ulem}

\usepackage{dsfont}
\usepackage{slashed}

\def\0#1#2{\frac{#1}{#2}}

\def\s0#1#2{\mbox{\small{$ \frac{#1}{#2} $}}}

\newcommand{\be}{\begin{eqnarray}}
\newcommand{\ee}{\end{eqnarray}}

\newcommand{\beq}{\begin{equation}}
\newcommand{\eeq}{\end{equation}}
\newcommand{\bea}{\begin{eqnarray}}
\newcommand{\eea}{\end{eqnarray}}

\usepackage{babel}
\makeatother
\begin{document}

\title{Magnetic-Field Induced Critical Endpoint}
\author{Stefan Rechenberger}
\affiliation{Institut f\"ur Theoretische Physik, Johann Wolfgang Goethe-Universit\"at, Max-von-Laue-Str. 1, D-60438 Frankfurt am Main, Germany}

\begin{abstract}

The phase diagram of strong interaction matter is analyzed utilizing the Nambu--Jona-Lasinio model. Special emphasis is placed on its dependence on an external magnetic field and isospin chemical potential. Using flavor mixing induced by instanton effects the influence of isospin breaking due to the magnetic field and the isospin chemical potential is compared. It is found that at low temperatures and large quark chemical potential the magnetic field, depending on its strength, induces a new critical endpoint or a triple point.
\end{abstract}

\maketitle

%
\section{Introduction}

It is known that strong magnetic fields of the order $m_\pi^2\approx 10^{18}Gauss$ are produced in off-central heavy ion collisions \cite{Kharzeev:2007jp, Skokov:2009qp, Bzdak:2011yy}. Furthermore, strong magnetic fields up to $10^{15}Gauss$ might appear on the surface of magnetars, a special type of neutron star \cite{Duncan:1992hi, Thompson:1993hn}. In the core of magnetars the magnetic field might even be stronger. Therefore, it is important to understand the influence of strong magnetic fields on strong interaction matter. Various interesting effects within heavy ion collisions as e.g. the chiral magnetic effect \cite{Kharzeev:2007jp, Fukushima:2008xe} are already known. For a recent review see \cite{Hattori:2016emy}.

The influence on the phase diagram of quantum chromodynamics (QCD) is versatile as well as reviewed in \cite{Kharzeev:2012ph, Andersen:2014xxa, Miransky:2015ava}. At vanishing quark chemical potential first lattice calculations have investigated the magnetic field dependence of the chiral critical temperature and found that it increases \cite{D'Elia:2010nq}. This is known as magnetic catalysis. More realistic calculations then showed that the chiral critical temperature actually decreases with the magnetic field \cite{Bali:2011qj}. This effect is known as inverse magnetic catalysis, anti catalysis or magnetic inhibition. Further simulations later showed evidence for a mixture of both behaviors \cite{Ilgenfritz:2013ara} which became known as delayed magnetic catalysis. Triggered by these results various attempts have been made to explain these intriguing effects \cite{Fukushima:2012kc, Kojo:2014gha, Braun:2014fua, Mueller:2015fka, Farias:2014eca, Ayala:2015bgv, Hattori:2015aki, Evans:2016jzo, Mao:2016fha, Mao:2016lsr, Tawfik:2016gye, Mamo:2015dea, Ayala:2014gwa, Ferreira:2014kpa, Gursoy:2016ofp}. At small temperatures and asymptotically large quark chemical potential the color superconducting phase is well established, see \cite{Alford:2007xm} for a review and \cite{Ferrer:2005vd, Ferrer:2006ie, Noronha:2007wg, Fukushima:2007fc} for the impact of a magnetic field. Approaching quark chemical potentials of the order of the constituent quark mass the situation is not that clear. The current understanding is that an inhomogeneous phase is present in this region of the phase diagram \cite{Bringoltz:2006pz, Bringoltz:2009ym, Nickel:2009wj, Kojo:2009ha, Carignano:2010ac, Carignano:2014jla, Buballa:2014tba, Lee:2015bva, Heinz:2015lua, Braun:2015fva}. The influence of a magnetic field on these phases was discussed in \cite{Frolov:2010wn, Ferrer:2012zq, Tatsumi:2014wka, Yoshiike:2015tha, Yoshiike:2015wud, Nishiyama:2015fba, Kashiwa:2015nya, Abuki:2016zpv, Cao:2016fby}.

Notably, quarks with different flavors couple differently to the magnetic field due to the corresponding charges. The up quark for example has charge $q_u=2/3\,e$ while the down quark has charge $q_d=-1/3\,e$ with $e$ being the elementary charge. Thus, the magnetic field explicitly breaks isospin symmetry. A second important source of isospin breaking is an explicit imbalance between up and down quarks described by a finite isospin chemical potential. This is relevant in the context of compact stars.

Since first principle lattice calculations are restricted to small values of the quark chemical potential our current knowledge of the biggest part of the phase diagram is based on low energy effective models for QCD as e.g. quark-meson models or linear sigma models. In this work I will discuss the dynamical chiral symmetry breaking within the Nambu--Jona-Lasinio (NJL) model with two flavors subject to an external magnetic field for varying temperature, quark chemical potential and isospin chemical potential. Furthermore, instanton effects are included into the analysis by introducing a 't~Hooft determinant term. In the present two-flavor scenario the latter can be translated into a new four-quark coupling of the NJL model. This term is introducing the $\mathrm{U_A}(1)$ breaking related to the axial anomaly. In most investigations the original NJL Lagrangian \cite{Nambu:1961tp, Nambu:1961fr} is used. In other words, the instanton interaction strength is chosen equal to the coupling strength of the $\mathrm{U_A}(1)$-symmetric four-Fermi interaction term. Consequently, the up and down quarks are mixed in a specific way. As a result, the corresponding condensates of the quark flavors coincide (see e.g. \cite{Costa:2013zca}). A different choice would be the $\mathrm{U_A}(1)$-symmetric scenario where the quark flavors decouple completely. In this setup the isospin breaking effect of the magnetic field or the isospin chemical potential results in a maximal split of the phase transitions corresponding to the up- and down-quark condensate respectively.

In \cite{Frank:2003ve} the occurrence of distinct phase transitions for the up and down quarks has been discussed in the context of non-zero isospin chemical potential. In this work I will compare this effect to the one caused by an external magnetic field as it was discussed in \cite{Ebert:1999ht, Boomsma_2010} at vanishing temperature and finite quark chemical potential. While extending this discussion to finite temperature I will discuss the fate of the new phase transitions which are induced by the magnetic field and correspond to a different amount of occupied Landau levels \cite{Ebert:1999ht, Boomsma_2010}.

This paper is organized as follows. The model under investigation is introduced in section \ref{sec:model}. After fixing the notation the mean-field approximation of the effective potential is derived subject to temperature, external magnetic field, quark chemical potential and isospin chemical potential. The numerical analysis of the phase structure is then comprised in section \ref{sec:ps} before concluding at the end.

\section{Model}\label{sec:model}
\subsection{Lagrangian and Symmetries}

In order to specify the notation I detail the NJL model by following the references \cite{Frank:2003ve, Boomsma_2010}. The Lagrangian reads
\begin{equation} \label{eq:NJLlagrangian}
\mathcal L_\mathrm{NJL} = \mathcal L_0 + \mathcal L_1 + \mathcal L_2
\end{equation}
with
\begin{eqnarray}
\mathcal L_0 &=& \bar\psi\left[ \imath\slashed\partial - m_\mathrm f - e_\mathrm f{\mathcal A}\!\!\!\slash{} 
- \gamma_0\mu_q - \gamma_0\tau_3\frac{\mu_\mathrm I}{2} \right]\psi \, , \nonumber \\
\mathcal L_1 &=& G_1\left[ (\bar\psi\psi)^2 + (\bar\psi\tau_i\psi)^2 - (\bar\psi\gamma_5\psi)^2 - (\bar\psi\tau_i\gamma_5\psi)^2 \right] , \nonumber \\
\mathcal L_2 &=& G_2\left[ (\bar\psi\psi)^2 - (\bar\psi\tau_i\psi)^2 + (\bar\psi\gamma_5\psi)^2 - (\bar\psi\tau_i\gamma_5\psi)^2 \right] \! .
\end{eqnarray}
Here, $\gamma_5$ denotes the fifth gamma matrix and the $\tau_i (i=1,2,3)$ are the Pauli matrices. Furthermore, $m_\mathrm f = \frac{m_u+m_d}{2}\mathbbm 1 + \frac{m_u-m_d}{2}\tau_3$ and $e_\mathrm f = \frac{e}{6}\mathbbm 1 + \frac{e}{2}\tau_3$ are the current quark mass matrix and the charge matrix in flavor space resulting in the masses and charges $m_u$ and $2/3 \, e$ for the up and $m_d$ and $-1/3 \, e$ for the down quark. The magnetic field is considered to be a constant background field with $\mathcal A_\mu = (0,0,B x_1,0)$. Finally, $\mu_q$ and $\mu_\mathrm I$ are the quark chemical potential and the isospin chemical potential respectively. They are related to the chemical potential for the up and down quark via $\mu_q = \frac{1}{2}(\mu_u + \mu_d)$ and $\mu_\mathrm I = (\mu_u - \mu_d)$. The interaction terms $\mathcal L_1$ and $\mathcal L_2$ comprise the attractive quark self interaction where $\mathcal L_2$ is the 't Hooft determinant term describing the instanton effects.

For vanishing current quark masses, magnetic field, isospin chemical potential and instanton effects, the Lagrangian \eqref{eq:NJLlagrangian} is symmetric under $\mathrm{SU_c}(3) \times \mathrm{SU_V}(2) \times \mathrm{SU_A}(2) \times \mathrm{U_V}(1) \times \mathrm{U_A}(1)$. While the instanton interaction is responsible for the axial anomaly and thus breaks the $\mathrm{U_A}(1)$ symmetry, equal current quark masses break explicitly the $\mathrm{SU_A}(2)$ symmetry. This results in $\mathrm{SU_c}(3) \times \mathrm{SU_V}(2) \times \mathrm{U_V}(1)$ symmetry reminiscent of QCD. Finally the inclusion of the magnetic field and the isospin chemical potential as well as a discrepancy between the current quark masses breaks the $\mathrm{SU_V}(2)$ symmetry. Furthermore, the latter symmetry is subject to spontaneous chiral symmetry breaking.

Performing a Hubbard-Stratonovich transformation one can trade the four-quark interactions for quark-meson interactions. Here I do not investigate the possibility of color superconducting phases or pion condensation and thus consider only condensates of the up and down quarks, $\langle\bar uu\rangle$ and $\langle\bar dd\rangle$, which might differ due to isospin breaking effects. The resulting bosonized Lagrangian in mean-field approximation reads
\begin{eqnarray}\label{eq:bosonizedL}
\mathcal L_\mathrm B &=& \mathcal L_0 -(G_1+G_2)\langle\bar\psi\psi\rangle^2 - (G_1-G_2)\langle\bar\psi\tau_3\psi\rangle^2 \nonumber \\
&&+ \bar\psi\left[ 2(G_1+G_2)\langle\bar\psi\psi\rangle + 2(G_1-G_2)\tau_3\langle\bar\psi\tau_3\psi\rangle \right]\psi \nonumber \\
&=& - 2G_1\left( \langle\bar uu\rangle^2 + \langle\bar dd\rangle^2 \right) - 4G_2 \langle\bar uu\rangle\langle\bar dd\rangle \nonumber \\
&& + \bar\psi\left[ \imath\slashed\partial - M_\mathrm f - e_\mathrm f{\mathcal A}\!\!\!\slash{} 
- \gamma_0\mu_q - \gamma_0\tau_3\frac{\mu_\mathrm I}{2} \right]\psi
\end{eqnarray}
with the constituent quark mass matrix $M_\mathrm f = \frac{M_u+M_d}{2}\mathbbm 1 + \frac{M_u-M_d}{2}\tau_3$ and
\begin{eqnarray}\label{eq:constituentMasses}
M_u &=& m_u -4G_1\langle\bar uu\rangle - 4G_2\langle\bar dd\rangle \, , \nonumber \\
M_d &=& m_d -4G_2\langle\bar uu\rangle - 4G_1\langle\bar dd\rangle \, .
\end{eqnarray}
The expressions \eqref{eq:bosonizedL} and \eqref{eq:constituentMasses} show that for $G_1=G_2$ and $m_u=m_d$ (as it is used in most NJL model investigations) the Lagrangian depends on the sum of the condensates $\langle\bar uu\rangle+\langle\bar dd\rangle$ and thus even for finite isospin chemical potential or magnetic field the chiral phase transition lines for the different flavors cannot be disentangled.

\subsection{Effective Potential}

For the investigation of the chiral phase transition an expression for the effective potential is needed which can be minimized to find the up- and down-quark condensates. Since the Hubbard-Stratonovich transformed Lagrangian \eqref{eq:bosonizedL} is quadratic in the fermion fields the path integral of the generating functional can be performed and the resulting effective potential within the imaginary time formalism and in the vacuum limit reads
\begin{eqnarray}\label{eq:vacEffectivePot}
\Omega &=& 2G_1\left( \langle\bar uu\rangle^2 + \langle\bar dd\rangle^2 \right) + 4G_2 \langle\bar uu\rangle\langle\bar dd\rangle \nonumber \\
&& - \mathrm{Tr}\ln\left[ \imath\gamma_0p_0 + \gamma_ip_i - M_\mathrm f \right] \nonumber \\
&=& 2G_1\left( \langle\bar uu\rangle^2 + \langle\bar dd\rangle^2 \right) + 4G_2 \langle\bar uu\rangle\langle\bar dd\rangle \nonumber \\
&& - 6 \sum_{f=u,d}\int\!\!\!\frac{\mathrm d^3\vec p}{(2\pi)^3}E_f \, .
\end{eqnarray}
Here the flavor dependent energy is defined as $E_f = \sqrt{\vec p^2 + M_f^2}$. Derivatives with respect to the condensates result in the vacuum gap equations
\begin{equation}\label{eq:vacGapEq}
\langle\bar ff\rangle = - 6\int\!\!\!\frac{\mathrm d^3\vec p}{(2\pi)^3} \frac{M_f}{E_f} \, , \qquad f=u,d \, .
\end{equation}
In order to keep the instanton interaction strength free to be chosen by hand four parameters have to be fixed: the current quark masses, a three-dimensional momentum cutoff and one coupling. Following \cite{Frank:2003ve, Boomsma_2010} a good choice is $m_u = m_d = m = 6 \,\mathrm{MeV}$, the cutoff $\Lambda = 590 \, \mathrm{MeV}$ and $G_0\Lambda^2 = 2.435$. The new coupling $G_0$ is related to $G_1$ and $G_2$ via
\begin{equation}
G_1 = (1-c) \, G_0 \, , \qquad G_2 = c \, G_0
\end{equation}
with the parameter $c$ to be chosen by hand. The choice $c=1/2$ results in the original NJL model \cite{Nambu:1961tp, Nambu:1961fr} and $M_u=M_d=M$ as well as $\langle\bar uu\rangle = \langle\bar dd\rangle = \frac{1}{2}\langle\bar qq\rangle$. The gap equations \eqref{eq:vacGapEq}, the Gel-Mann--Oakes--Renner relation \cite{GellMann:1968rz}
\begin{equation}
f_\pi^2m_\pi^2 = -m\langle\bar qq\rangle
\end{equation}
and the relation (as derived e.g. in \cite{Klevansky:1992qe}) for the pion decay constant
\begin{equation}
f_\pi^2 = 3 M^2 \int\!\!\!\frac{\mathrm d^3\vec p}{(2\pi)^3} \left( \vec p^2 + M^2 \right)^{-\frac{3}{2}}
\end{equation}
together with the above choice of parameters give reasonable values for the pion decay constant $f_\pi = 92.6 \, \mathrm{MeV}$ the pion mass $m_\pi = 140.2 \, \mathrm{MeV}$ and the condensate $\langle\bar uu\rangle = -(241.5 \, \mathrm{MeV})^3$.

Finite temperature and chemical potentials can be incorporated by using within the trace of \eqref{eq:vacEffectivePot} the replacement rules
\begin{equation}
p_0\rightarrow \nu_n+\imath\mu_f \qquad
\text{and} \qquad
\int\!\!\frac{\mathrm dp_0}{2\pi} \rightarrow T\sum_n
\end{equation}
with fermionic Matsubara frequencies $\nu_n = 2\pi T(n+\frac{1}{2})$. The magnetic field in turn can be considered by using Schwinger's proper time method \cite{Schwinger:1951nm, Ebert:1999ht} or following \cite{Menezes:2008qt} by summing over Landau levels after replacing
\begin{align}
p_1^2+p_2^2 \rightarrow |e_fB|(2n+1)-\imath e_fB\gamma_1\gamma_2 \nonumber \\
\text{and} \qquad
\int\!\!\frac{\mathrm dp_1}{2\pi}\int\!\!\frac{\mathrm dp_2}{2\pi} \rightarrow \frac{|e_fB|}{2\pi}\sum_k \, .
\end{align}
The resulting effective potential contains a sum over spins $s$ and  reads\footnote{An irrelevant electromagnetic contribution coming from $(F_{\mu\nu})^2$ is suppressed.}
\begin{widetext}
\begin{eqnarray}\label{eq:efPotunint}
\Omega(T,B,\mu_u,\mu_d)
&=& 2G_0\left( \langle\bar uu\rangle^2 + \langle\bar dd\rangle^2 -c\left( \langle\bar uu\rangle-\langle\bar dd\rangle\right)^2\right) \nonumber \\
&&- \sum_{f=u,d} \frac{3|e_fB|}{2\pi}T\sum_{n=-\infty}^\infty\sum_{k=0}^\infty\sum_{s=\pm 1} \int\!\!\frac{\mathrm dp_3}{2\pi} \ln
\left[(\nu_n+\imath\mu_f)^2 + p_3^2 + |e_fB|(2k+1+s) - M_f^2 \right] \nonumber \\
&=& 2G_0\left( \langle\bar uu\rangle^2 + \langle\bar dd\rangle^2 -c\left( \langle\bar uu\rangle-\langle\bar dd\rangle\right)^2\right) - \sum_{f=u,d} \frac{3|e_fB|}{2\pi}\sum_{k=0}^\infty(2-\delta_{k,0}) \int\!\!\frac{\mathrm dp_3}{2\pi} E_f \\
&&- \sum_{f=u,d} \frac{3|e_fB|}{2\pi}T\sum_{k=0}^\infty(2-\delta_{k,0}) \int\!\!\frac{\mathrm dp_3}{2\pi} \left\{
\ln\left[1+\exp\left( -(E_f - \mu_f)/T \right) \right]
+ \ln\left[1+\exp\left( -(E_f + \mu_f)/T \right) \right] \right\} \, . \nonumber
\end{eqnarray}
\end{widetext}
In the second step the sum over Matsubara frequencies has been performed and the flavor dependent energy was defined as $E_f^2 = p_3^2 + 2|e_fB|k + M_f^2$. Along the lines of \cite{Menezes:2008qt} the last part of the third line in \eqref{eq:efPotunint} can be evaluated analytically and the final expression for the effective potential can be split into three parts:
\begin{equation}\label{eq:effectivePot}
\Omega(T,B,\mu_u,\mu_d) = \Omega_0 + \Omega_1(B) + \Omega_2(T,B,\mu_u,\mu_d)
\end{equation}
with
\begin{widetext}
\begin{align}
\Omega_0 &= 2G_0\left( \langle\bar uu\rangle^2 + \langle\bar dd\rangle^2 -c\left( \langle\bar uu\rangle-\langle\bar dd\rangle\right)^2\right) - 6\sum_{f=u,d}\int\!\!\frac{\mathrm d^3\vec p}{(2\pi)^3}\sqrt{\vec p^2+M_f^2} \, , \nonumber \\
\Omega_1 &= - \sum_{f=u,d}\frac{3(e_fB)^2}{2\pi^2}\left[ \frac{x_f^2}{4} - \frac{x_f^2-x_f}{2}\ln x_f + \zeta'(-1,x_f) \right] \, , \nonumber \\
\Omega_2 &= - \sum_{f=u,d} \frac{3|e_fB|}{2\pi}T\sum_{k=0}^\infty(2-\delta_{k,0}) \int\!\!\frac{\mathrm dp_3}{2\pi} \left\{
\ln\left[1+\exp\left( -(E_f - \mu_f)/T \right) \right]
+ \ln\left[1+\exp\left( -(E_f + \mu_f)/T \right) \right] \right\} \, .
\end{align}
\end{widetext}
Here, $\zeta$ is the Hurwitz zeta function and $\zeta'(-1,x_f)=\frac{\mathrm d}{\mathrm dz}\zeta(z,x_f)|_{z=-1}$. The shortcut $x_f$ is defined via $M_f^2 = 2|e_fB|x_f$ and the divergent vacuum contribution $\Omega_0$ contains a three-dimensional momentum cutoff.

\section{Chiral Phase Structure}\label{sec:ps}
In this section the above derived effective potential is analyzed depending on the external parameters $T,B,\mu_u$ and $\mu_d$ with special emphasis on the isospin breaking effects of the magnetic field and the isospin chemical potential. Afterwards, a new phase transition, induced by the magnetic field, in the cold and dense part of the phase diagram is discussed.
\subsection{Chiral Phase Structure and Isospin Breaking}
As mentioned in the previous section the effective potential contains the external parameters $T, B, \mu_q$ and $\mu_\mathrm I$, the coupling $G_0$ and the three-dimensional momentum cutoff $\Lambda$ as model parameters and finally the parameter $c$ specifying the 't Hooft determinant interaction strength. The latter is to be chosen by hand. Notably, this parameter can be estimated in the context of the three-flavor model including a strange-quark condensate to be of the order $c\simeq 0.2$ \cite{Frank:2003ve}. The goal of this work is a comparison of isospin breaking effects of the isospin chemical potential and the external magnetic field. Thus, the presentation of the numerical results concentrates on those obtained for $c=0$ yielding the largest split of the phase transition lines. This allows a clear presentation of the effects without changing the qualitative behavior.

\begin{figure}
\includegraphics[width=0.4\textwidth]{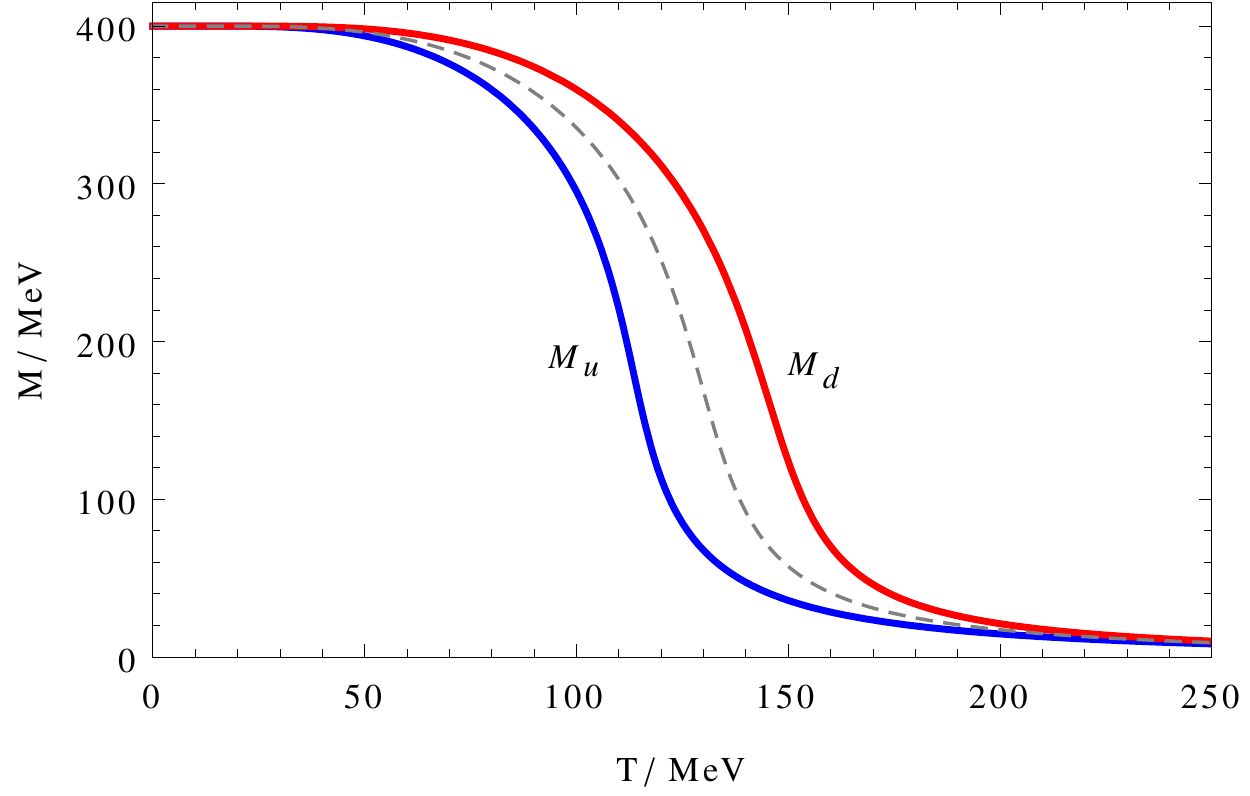}
\includegraphics[width=0.4\textwidth]{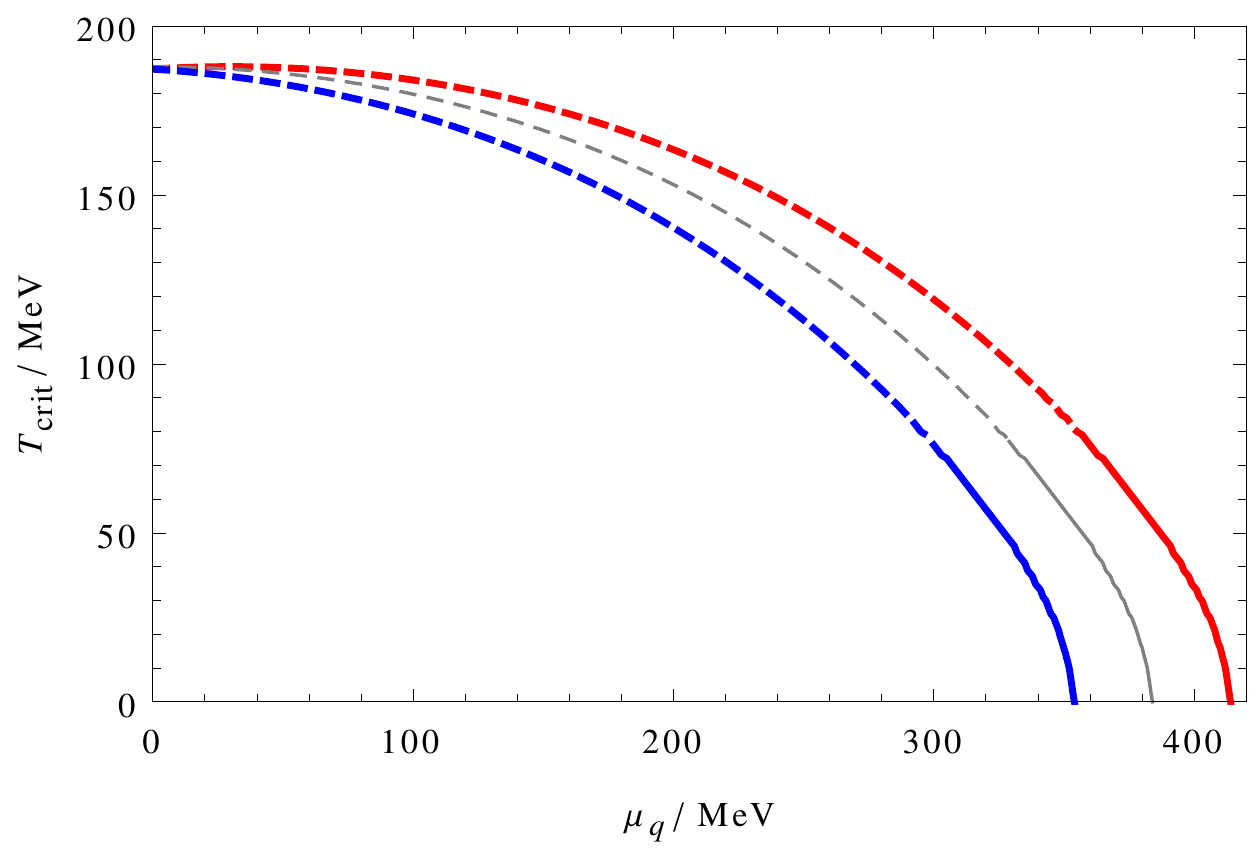}
\caption{Upper plot: Consitutent quark masses $M_u$ and $M_d$ depending on the temperature for $B=0, \mu_q=250\,\mathrm{MeV}$ and $\mu_\mathrm I=60\,\mathrm{MeV}$. For comparison the result for $c=1/2$ is depicted in gray.
Lower plot: Dashed crossover lines and solid first order lines for the up-quark condensate (blue) and the down-quark condensate (red) for $B=0$ and $\mu_\mathrm I=60\,\mathrm{MeV}$. For comparison the result for $\mu_\mathrm I = 0$ is depicted in gray.}
\label{fig:muI}
\end{figure}

To start with we use vanishing external magnetic field and $\mu_\mathrm I = 60\,\mathrm{MeV}$. At lager isospin chemical potentials pion superfluidity, as studied e.g. in \cite{Cao:2015xja}, should be included in the analysis. At $\mu_\mathrm I = 60\,\mathrm{MeV}$ the resulting temperature dependence of the constituent quark masses is depicted in the upper plot of Figure \ref{fig:muI} for $\mu_q = 250\,\mathrm{MeV}$.
The typical crossover behavior is obtained for both constituent quark masses. However, both curves are clearly separated and distributed symmetrically around the gray line. The latter one corresponds to the result for $c=1/2$ not allowing for a distinct condensation of the different flavors as explained above.

Assigning a crossover temperature to both curves via the corresponding turning points and varying the quark chemical potential results in the dashed curves of the lower plot in Figure \ref{fig:muI}. Since the flavors decouple completely for $c=0$ and the chemical potentials for the flavors differ only by a sign ($\mu_u=-\mu_d$) for $\mu_q=0$ the crossover lines agree at vanishing quark chemical potential with each other and with the $\mu_\mathrm I=0$ result depicted in gray. For larger values of the quark chemical potential the crossovers turn into first order transitions (depicted as solid lines in Figure \ref{fig:muI}) and corresponding critical endpoints. The critical endpoints have identical temperatures but varying values for the quark chemical potential. For $T=0$ the separation of the first order transitions is given by $\mu_\mathrm I = 60\,\mathrm{MeV}$. This agrees well with \cite{Klein:2003fy, Toublan:2003tt, Frank:2003ve}. Note however, that this observation is restricted to the case $c=0$ where the flavors decouple completely. Larger values of $c$ decrease the separation between the transition lines. For more realistic values $c\simeq 0.2$ the first order lines of the up- and down-quark condensates agree and the difference for the crossover is reduced drastically (see also \cite{Frank:2003ve, He:2005sp}).

\begin{figure}[b]
\includegraphics[width=0.4\textwidth]{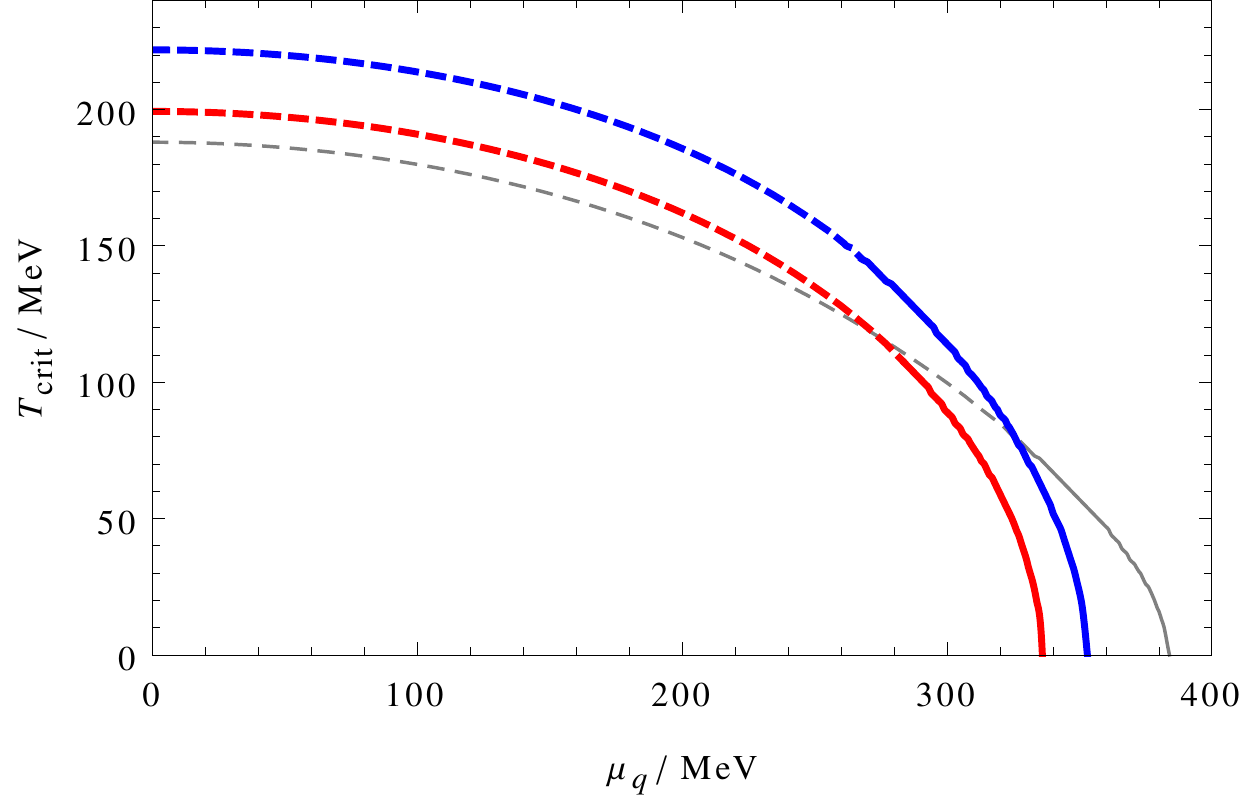}
\caption{Dashed crossover lines and solid first order lines for the up-quark condensate (blue) and the down-quark condensate (red) for $|eB|=20\,m_\pi^2$ and $\mu_\mathrm I=0$. For comparison the result for $|eB| = 0$ is depicted in gray.}
\label{fig:B}
\end{figure}

A similar split of the critical temperatures for the up- and down-quark condensates is expected to occur when introducing a magnetic field due to the different charges $q_u = 2/3\,e$ and $q_d = -1/3\,e$ coupling the quarks to the external field. In the following academically strong magnetic fields $|eB|=20\,m_\pi^2$ as well as $c=0$ shall be discussed for presentational reasons. For weaker fields the resulting split of the transition lines decreases.
Assigning again crossover temperatures to the turning points of the temperature dependences of the constituent quark masses the resulting phase diagram is depicted in Figure \ref{fig:B}.
The first striking feature is the increase of the chiral critical temperature of the up-quark condensate (blue) and the down-quark condensate (red) at vanishing quark chemical potential in comparison to the result for vanishing external field (gray). This is the magnetic catalysis effect which is known to be an artifact of the simplified model analysis considered here \cite{Bali:2011qj, Ilgenfritz:2013ara, Fukushima:2012kc, Kojo:2014gha, Braun:2014fua, Mueller:2015fka, Hattori:2015aki, Evans:2016jzo, Mao:2016fha, Mao:2016lsr, Tawfik:2016gye, Mamo:2015dea, Ferreira:2014kpa, Gursoy:2016ofp}. This effect is stronger for the up quark due to the flavor dependent charges.
For larger values of the quark chemical potential the crossovers turn into first order transitions again which results in critical endpoints. For even larger values of $\mu_q$ the magnetic catalysis turns into an inverse catalysis i.e. the chiral critical temperature decreases with the magnetic field. This effect is more pronounced for the down quark and thus the critical quark chemical potential of the down quark at vanishing temperature is smaller than the one for the up quark.
Note that for very strong magnetic fields this decrease is turned into an increase again \cite{Inagaki:2003yi, Inagaki:2004ih}.
Changing the instanton effects i.e. increasing $c$ reduces the split of the two separate transition lines and ends at $c=1/2$ with one single line in the middle of the blue and the red one in Figure \ref{fig:B}. This resulting line then shows magnetic catalysis at small $\mu_q$ as well as inverse catalysis at large $\mu_q$.

\begin{figure}
\includegraphics[width=0.4\textwidth]{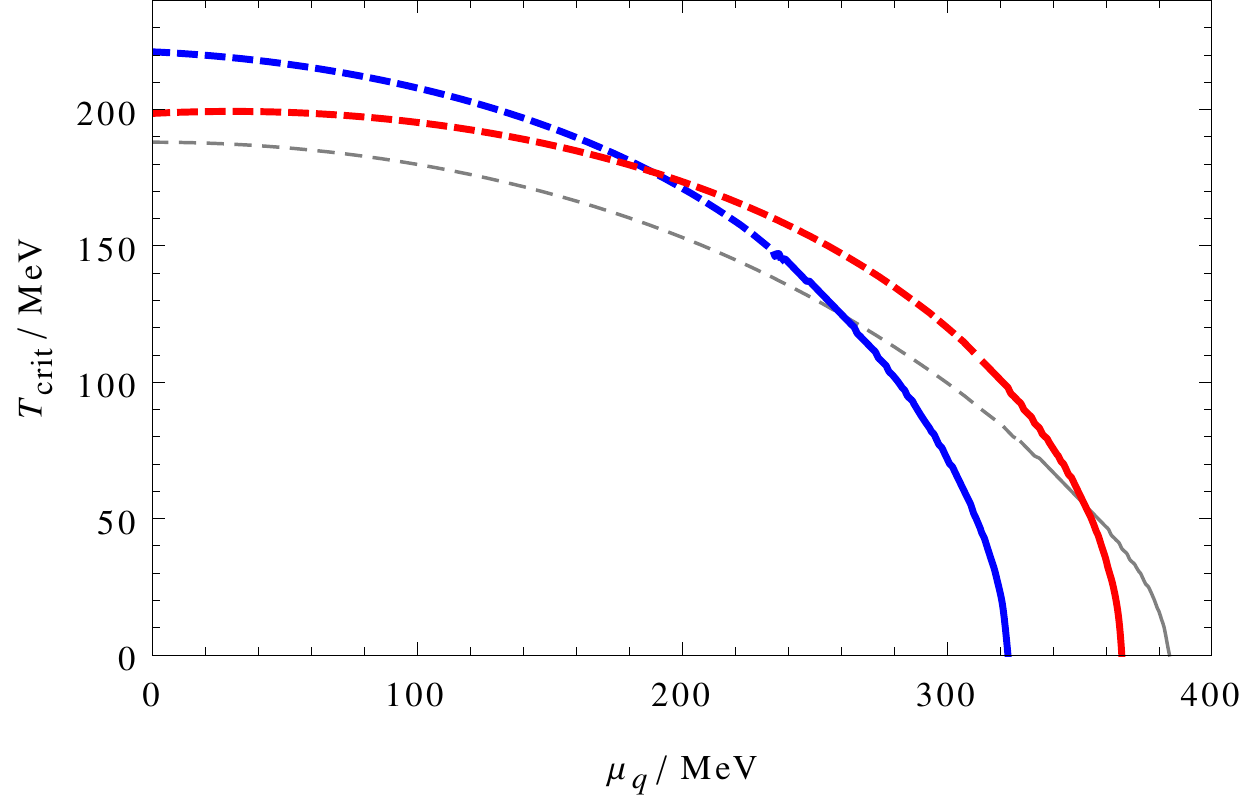}
\caption{Dashed crossover lines and solid first order lines for the up-quark condensate (blue) and the down-quark condensate (red) for $|eB|=20\,m_\pi^2$ and $\mu_\mathrm I=60\,\mathrm{MeV}$. For comparison the result for $|eB| = 0$ and $\mu_\mathrm I=0$ is depicted in gray.}
\label{fig:BmuI}
\end{figure}

A direct comparison of the flavor dependent separation of the transition lines caused by isospin chemical potential in Figure \ref{fig:muI} and the one caused by the magnetic field in Figure \ref{fig:B} shows that the former split is symmetric around the $\mu_\mathrm I = 0$ result while the latter is asymmetric around $B=0$. Finally the combined effect for $|eB|=20\,m_\pi^2$ and $\mu_\mathrm I=60\,\mathrm{MeV}$ and again $c=0$ is depicted in Figure \ref{fig:BmuI}.
Since the inclusion of $\mu_\mathrm I$ does not influence the crossover lines at $\mu_q=0$ as discussed in Figure \ref{fig:muI} the corresponding results agree in Figure \ref{fig:B} and Figure \ref{fig:BmuI}. For increasing $\mu_q$ however $\mu_\mathrm I$ has a considerable influence on the phase structure. It tends to increase the chiral critical temperature for the down-quark condensate and decrease the one for the up-quark condensate. This results in a point around $\mu_q \simeq 200\,\mathrm{MeV}$ where both crossover temperatures agree. Increasing $\mu_q$ further splits again the two crossover lines which turn into first order lines at the corresponding critical endpoints. The interchanged role of the up and down quark flavor for $T=0$ in comparison to Figure \ref{fig:B} is caused not only by $\mu_\mathrm I$ but also by the fact that the inverse catalysis turns into a catalysis for very strong magnetic fields \citep{Inagaki:2003yi, Inagaki:2004ih}. The strength of the magnetic field at which this change from inverse catalysis to catalysis sets in is again flavor dependent and can cause the interchanged role of the flavors even without $\mu_\mathrm I$.

\subsection{Magnetic-Field Induced Critical Endpoint}

While analyzing the influence of an external magnetic field on the phase structure at vanishing temperatures new phases have been found \cite{Boomsma_2010, Ebert:1999ht}. They appear at sufficiently large quark chemical potential and correspond to dynamically broken chiral symmetry, non-zero nuclear density and a certain amount of filled Landau levels. If only the lowest Landau level is filled the new phase is called $C_0$ in the nomenclature of \cite{Boomsma_2010, Ebert:1999ht}. It is situated between the chirally broken and the restored phase and is bounded by first order phase transitions. The fate of this transition line at finite temperature shall be discussed in the following.

\begin{figure}[b]
\includegraphics[width=0.4\textwidth]{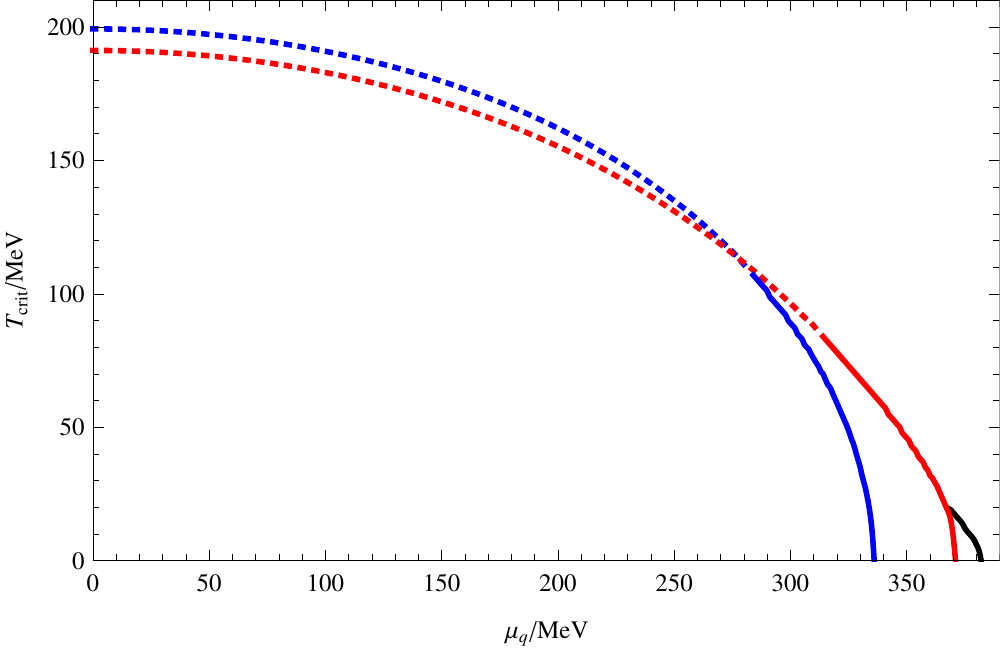}
\caption{Dashed crossover lines and solid first order lines for the up-quark condensate (blue) and the down-quark condensate (red) for $|eB|=10\,m_\pi^2$ and $\mu_\mathrm I=0$. The new phase $C_0$ for the down quark is bounded by the solid, black first order line.}
\label{fig:newPhase}
\end{figure}

\begin{figure}
\includegraphics[width=0.23\textwidth]{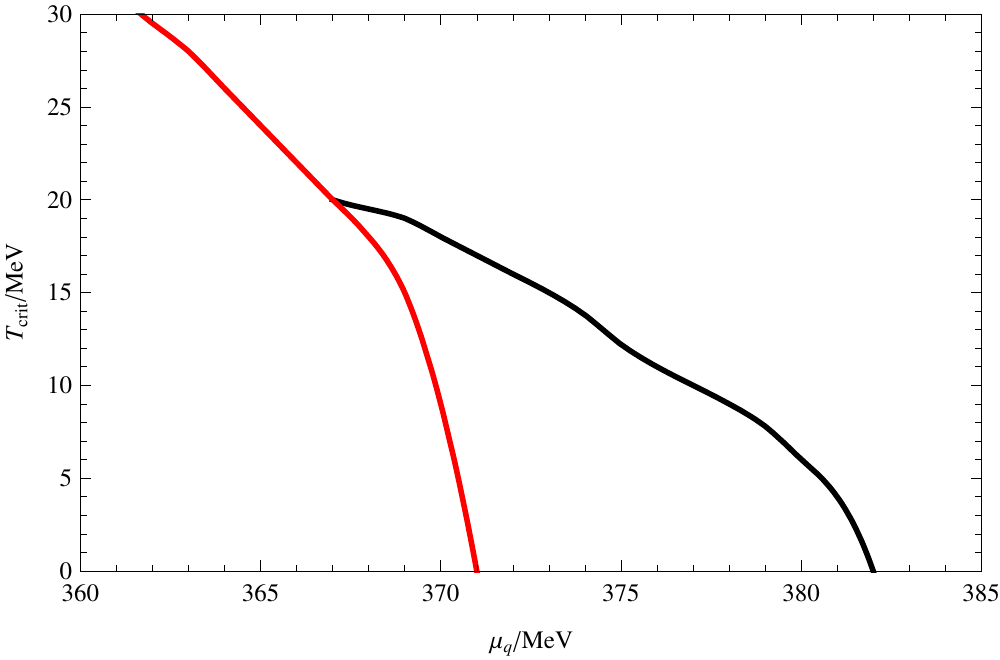}
\includegraphics[width=0.23\textwidth]{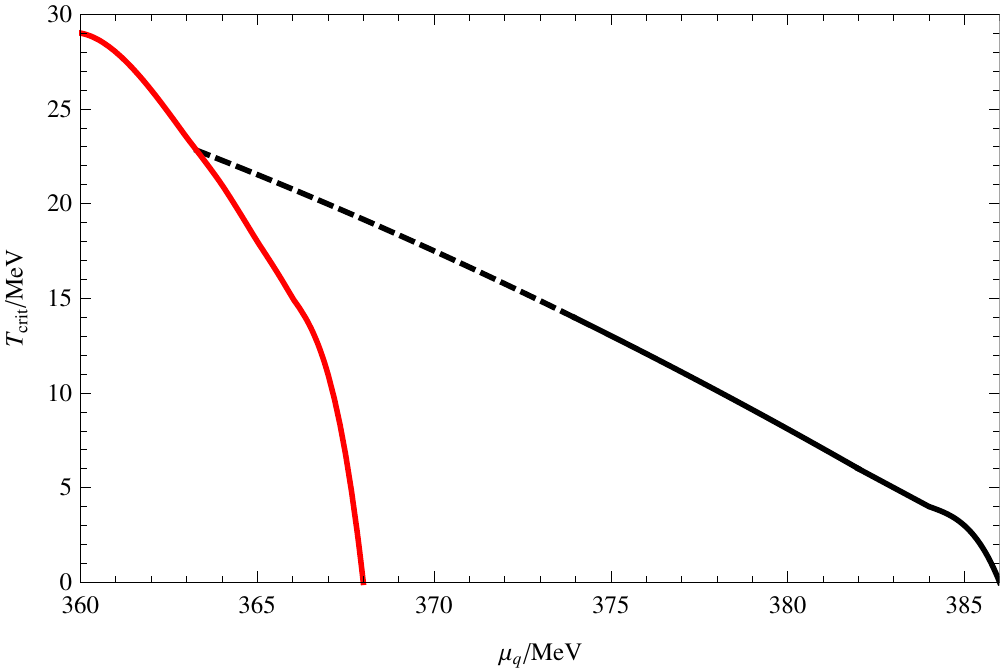}
\caption{Left plot: Zoomed version of Figure \ref{fig:newPhase} with $|eB|=10\,m_\pi^2$. The phase $C_0$ is enclosed by the red and black first order lines.
Right plot: Same plot, but with $|eB|=10.5\,m_\pi^2$. The solid, black first order line ends in a new critical endpoint.}
\label{fig:zoom}
\end{figure}

For simplicity and comparability to the results of \cite{Boomsma_2010} I restrict myself to the case of vanishing isospin chemical potential and complete decoupling of the flavors i.e. $c=0$. The corresponding phase diagram is depicted in Figure \ref{fig:newPhase} for $|eB|=10\,m_\pi^2$ and $\mu_\mathrm I=0$.
It shows the interchanged role of up- and down-quark condensate at small and large values of $\mu_q$ as in Figure \ref{fig:BmuI}. Since $\mu_\mathrm I=0$ this is caused by the magnetic catalysis at large values of $\mu_q$ and $B$ as discussed in the previous subsection.
The new phase $C_0$ with broken chiral symmetry, non-zero nuclear density and filled lowest Landau level appears at small temperatures and large values of $\mu_q$. It is separated from the chirally broken phase with zero nuclear density by the red, solid first order line. The black, solid first order line depicts the border to the phase with restored chiral symmetry (up to the explicit breaking due to quark masses). A zoomed version of that new phase is depicted in the left plot of Figure \ref{fig:zoom}.
The red line is the first order transition discriminating between zero and non-zero nuclear density to its left and right respectively. The black, solid line denotes a very weak first order transition bordering the $C_0$ phase to its left. The two first order transitions meet at a triple point.
Varying the strength of the external magnetic field  from $|eB|=10\,m_\pi^2$ to $|eB|=10.5\,m_\pi^2$ weakens the first order transition between the $C_0$ phase and the phase with restored chiral symmetry even further and for temperatures above $T\simeq 15\,\mathrm{MeV}$ turns it into a crossover. Therefore, a new critical endpoint induced by the magnetic field develops. A similar new phase with corresponding first-order or crossover lines can be found for the up-quark condensate. Notably, this appears at smaller values for the external magnetic field around $|eB|\simeq 5\, m_\pi^2$. Interestingly, the existence of this new critical endpoint appears to be stable against the inclusion of vector interactions \cite{Denke:2013gha}.

At this point some comments are in order. The values of the constituent up quark mass changes discontinuously at the corresponding first order transition. The constituent down quark mass in turn jumps at the first order transition corresponding to the down quark condensate. This rather simple picture is true only in the case of completely decoupled flavors $c=0$. However, for realistic values of the instanton effect, $c\simeq 0.2$, the flavors couple and the discontinuity of the constituent quark mass of one flavor causes a discontinuity in the second flavor's constituent quark mass. Another effect of $c>0$ is a smaller gap between the phase transition lines of the two flavors. In combination with the various first order transitions between the magnetically induced new phases with variable amount of filled Landau levels (for both flavors) causes many discontinuities in the $T$- or $\mu_q$ dependence of the order parameter. This makes it numerically very hard to disentangle them and assign a primary cause to each individual jump.

\section{Conclusions}\label{sec:conc}

It is known that strong magnetic fields are created in off-central heavy ion collisions \cite{Kharzeev:2007jp, Skokov:2009qp, Bzdak:2011yy} as well as in neutron stars \cite{Duncan:1992hi, Thompson:1993hn}. Therefore, a thorough understanding of the influence of magnetic fields on strong interaction matter is of utmost interest.
Many interesting effects have been discovered in the past \cite{Hattori:2016emy} and the aim of this work is to contribute to the understanding of the dependence of the QCD phase diagram on an external magnetic field. Special emphasis is placed on the isospin breaking due to the flavor dependent charge coupling the external fields to the quarks.
For this purpose the two-flavor NJL model is utilized.

To be more precise, the quarks of the NJL model are coupled to an external magnetic field and the system is considered subject to finite temperature, quark chemical potential and isospin chemical potential. In order to allow for a non-trivial coupling of the flavors instanton effect, effectively incorporated as a 't Hooft determinant term into the model's Lagrangian, are included in the analysis. Varying the corresponding parameter $c$ allows the discussion of the completely decoupled case for $c=0$ (vanishing 't Hooft determinant term) as well as the original NJL model \cite{Nambu:1961tp, Nambu:1961fr} for $c=1/2$ ('t Hooft determinant interaction strength equals four-Fermi interaction strength).
The effective potential, evaluated using the mean-field approximation, then allows for a discussion of dynamical chiral symmetry breaking depending on the four external parameters and the instanton effects.

Both, the external magnetic field and the isospin chemical potential introduce an asymmetry between the two flavors into the system, i.e. break isospin symmetry. This results in a flavor dependent condensation behavior and thus separate transition and crossover lines for the chiral phase transition of the up quark and the down quark. As shown in Figure \ref{fig:muI} this split appears symmetrically around the isospin symmetric result if it is caused by the isospin chemical potential (see also \cite{Frank:2003ve}). The external magnetic field on the other hand turns out to shift both phase transitions into the same direction (relative to the $B=0$ result) but in a different amount due to the different charges of the quark flavors as depicted in Figure \ref{fig:B} (see also \cite{Boomsma_2010}). The size of the gap between the two transition lines depends on the 't Hooft determinant interaction strength and is largest for completely decoupled flavors, i.e. $c=0$. For increasing instanton effects, corresponding to growing $c$, the size of the gap decreases. Varying, the magnetic field strength and the isospin chemical potential results in a variety of phases condensed and non-condensed up and down quarks. An example is given in Figure \ref{fig:BmuI}.

In case the system is subject to an external magnetic field at vanishing temperatures the transition from the phase with broken chiral symmetry and vanishing nuclear density to the phase with restored chiral symmetry and non-zero nuclear density takes placed in several steps. Between the latter two phases many phases with broken chiral symmetry and non-zero nuclear density are found which are separated by first order transitions \cite{Ebert:1999ht, Boomsma_2010}. While moving from the chirally broken to the restored phase the Landau levels of the quarks subject to the magnetic field are filled one after the other leading to these intermediate phases. Analyzing the temperature dependence of the transition line bounding the phase with filled lowest landau level results in a behavior depending on the strength of the magnetic field. One possibility is a change from a first order transition at vanishing temperature to a crossover behavior at larger temperatures and a corresponding magnetic-field induced critical endpoint as depicted in the right plot of Figure \ref{fig:zoom}. The other option is a persistent first order transition line merging with the first order transition line at the border to the vanishing-nuclear-density phase in a triple point as depicted in the left plot of Figure \ref{fig:zoom}. In total the phase structure of strong interaction matter subject to an external magnetic field and isospin chemical potential is very versatile especially at low temperatures and large quark chemical potentials.

\text{ }\newline
{\it Acknowledgments.--~} I thank Gabor Almasi, Jens Braun, Rob Pisarski, Juergen Schaffner-Bielich and Vladimir Skokov for many helpful discussions. Furthermore, I acknowledge the hospitality of the Brookhaven National Laboratory, where most of this work has been done. This work was supported by a postdoc fellowship of the German Academic Exchange Service (DAAD).


%
\bibliography{bib}

\end{document}